\documentclass[12pt]{JHEP3}

\usepackage{cite}
\usepackage{enumerate}
\usepackage{amsbsy}
\usepackage[dvips]{graphicx}
\usepackage{graphics}

\newcommand{\be}{\begin{equation}} \newcommand{\ee}{\end{equation}}
\newcommand{\bea}{\begin{eqnarray}} \newcommand{\eea}{\end{eqnarray}}

\newcommand{\re}[1]{(\ref{#1})}

\newcommand{\pat}{\partial}

\renewcommand{\sec}[1]{section \ref{#1}}
\newcommand{\fig}[1]{figure \ref{#1}}
\newcommand{\brt}[1]{[#1]}

\renewcommand{\a}{\alpha}
\renewcommand{\b}{\beta}

\newcommand{\GN}{G_{\mathrm{N}}}
\newcommand{\ha}{\frac{1}{2}}

\newcommand{\keq}{k_{\mathrm{eq}}}

\newcommand{\teq}{t_{\mathrm{eq}}}

\newcommand{\rmd}{\mathrm{d}}
\newcommand{\bz}{\bar{z}}

\newcommand{\nonum}{\\}
\newcommand {\etal} {et al.}

\newcommand{\rhodot}{\dot{\rho}}

\newcommand{\HH}{\frac{{\dot a}_\CD^2}{a_\CD^2}}

\newcommand{\av}[1]{\langle{#1}\rangle_\CD}
\newcommand{\sQ}{\mathcal{Q}_\CD}
\newcommand{\mQ}{\mathcal{Q}_\CM}
\newcommand{\eQ}{\mathcal{Q}_\CE}
\newcommand{\sR}{{^{(3)}R}}

\newcommand{\CD}{{\cal D}}

\newcommand{\CE}{{\cal E}}

\newcommand{\CM}{{\cal M}}

\newcommand{\average}[1]{\left\langle #1 \right\rangle_\CD}

\newcommand{\PRD}[1]{{\it Phys. Rev.} {\bf D#1}}

\newcommand{\PRL}[1]{{\it Phys. Rev. Lett.} {\bf #1}}

\newcommand{\NPB}[1]{{\it Nucl. Phys.} {\bf B#1}}
\newcommand{\PLA}[1]{{\it Phys. Lett.} {\bf A#1}}

\newcommand{\MNRAS}[1]{{\it Mon. Not. Roy. Astron. Soc.} {\bf #1}}
\newcommand{\APJ}[1]{{\it Astrophys. J.} {\bf #1}}

\newcommand{\JHEP}[1]{{\it JHEP} {\bf #1}}
\newcommand{\CQG}[1]{{\it Class. Quant. Grav.} {\bf #1}}
\newcommand{\GRG}[1]{{\it Gen. Rel. Grav.} {\bf #1}}
\newcommand{\AaA}[1]{{\it Astron. \& Astrophys.} {\bf #1}}
\newcommand{\PROG}[1]{{\it Prog. Theor. Phys.} {\bf #1}}

\newcommand{\IJMPD}[1]{{\it Int. J. Mod. Phys.} {\bf D#1}}

\title{Backreaction in late-time cosmology}

\author{Thomas Buchert
\\ Universit\'e de Lyon, Observatoire de Lyon, CNRS UMR 5574,
Centre de Recherche Astrophysique de Lyon: Universit\'e Lyon 1 and \'Ecole
Normale Sup\'erieure de Lyon \\
9 avenue Charles Andr\'e, F-69230 Saint-Genis-Laval, France \\
\email{buchert {\it at} obs {\it dot} univ-lyon1 {\it dot} fr}}

\author{Syksy R\"{a}s\"{a}nen
\\ University of Helsinki, Department of Physics \\
and Helsinki Institute of Physics \\
P.O. Box 64, FIN-00014 University of Helsinki, Finland \\
\email{syksy {\it dot} rasanen {\it at} iki {\it dot} fi}}

\abstract{
We review the effect of the formation of nonlinear
structures on the expansion rate, spatial curvature
and light propagation in the universe, focusing 
on the possibility that it could explain cosmological
observations without the introduction of dark energy or
modified gravity.
We concentrate on explaining the relevant physics and
highlighting open questions.
}

\preprint{HIP-2011-35/TH}

\keywords{general relativity, cosmology, dark energy, structure formation, Newtonian limit}

\begin{document}
  
\setcounter{tocdepth}{2}

\setcounter{secnumdepth}{3}

\vspace{-1 truecm}

\section{Introduction} \label{sec:intro}

\subsection{The role of structure formation in cosmology}

The formation of nonlinear structures from tiny almost
Gaussian perturbations by gravitational instability is
a central topic in cosmology.
Most studies are done either in general relativity (GR) using
linear perturbation theory around the locally homogeneous
and isotropic Friedmann-Lema\^{\i}tre-Robertson-Walker (FLRW) model
or in Newtonian gravity, where extensive numerical computations
have been done with N-body simulations,
preceded and accompanied by analytical work,
to access the nonlinear regime of structure formation.
The formation of structures is sensitive to the expansion of
the universe, so the observed large-scale structure is a valuable
probe of cosmological evolution. There is also another side to the
matter: perturbations have a significant effect on the
expansion rate once they enter the nonlinear regime.
In standard linear theory, the effect vanishes on average
by construction. In Newtonian gravity, this turns out
to be true also in the non-perturbative regime (see \sec{sec:newton}).
However, this result does not carry over to non-perturbative GR, where
the geometry is dynamical. Therefore, there could be
a large effect on the average expansion rate, the evolution of
spatial curvature and light propagation.

The effect of deviations from exact homogeneity and isotropy on the
average expansion is known as {\it backreaction}. The effect was first
studied in 1962 \cite{Shirokov:1962}, and an in-depth
discussion was given 20 years later by George Ellis \cite{fitting}.
Ellis introduced the term {\it fitting problem} to describe the
issue of finding a model that best describes the average
of the real universe with its complex structures.
The problems laid out by Ellis and collaborators did not become
mainstream concerns in cosmology at the time, which may partly be
due to the complexity of the issues. Another likely reason is
that cosmological observations were not precise enough to indicate any
tension with a homogeneous and isotropic cosmological
model based on ordinary GR with ordinary (visible and dark) matter.
As more precise data has revealed that such a model disagrees
with observations, it has become topical to study
backreaction in more detail to see whether it could account
for this failure
\cite{Buchert:2000, Wetterich:2001, Schwarz:2002, Rasanen}.
We review backreaction in classical GR
due to the formation of nonlinear structures at late times,
with a focus on the possibility that it could explain this discrepancy.
We do not aim to give a historical overview of the literature, but
rather try to explain the relevant physics and highlight open
questions.
There are a number of reviews with comprehensive lists of references
\cite{Rasanen:2006b,buchert:review,ellisbuchert,Ellis:2011a,buchertFOCUS, Rasanen:2011a, Clarkson:2011b}.
We concentrate on structures that are known to exist and
that are expected in the standard structure formation scenario
based on usual models of inflation.
In particular, we do not discuss models
with Gpc-scale structures where, for instance, the observer
is located near the centre of a large void \cite{voidmodels}.
We do not discuss the backreaction of classical or
quantum perturbations during inflation or reheating \cite{quantum}.
We also do not cover schemes that introduce some
extra structures to GR, which may be needed if the
`fitting problem' is to be solved for tensorial
objects such as the metric \cite{nonscalar}.

In section 1 we briefly review some relevant cosmological
observations and discuss assumptions underlying the FLRW model.
In section 2 we put into perspective the equations that govern the average
expansion, explain the meaning of average acceleration
and discuss the difference between GR and Newtonian gravity.
In section 3 we discuss estimates of backreaction in perturbation
theory and in non-perturbative models.
In section 4 we consider light propagation and the relation
of spatially averaged quantities to observables, as well as
observational signatures of backreaction.
We conclude in section 5 with a summary.

\subsection{Cosmological observations} \label{sec:obs}

A linearly perturbed homogeneous and isotropic cosmological model
with ordinary matter (i.e. matter with non-negative pressure)
and ordinary gravity (based on the four-dimensional Einstein-Hilbert action)
can explain all observations of the early universe from
Big Bang Nucleosynthesis onwards. The simplest such model,
known as the Standard Cold Dark Matter (SCDM) model, is
matter-dominated at late times and has flat spatial sections.
The SCDM model can also account for late-time observations at the accuracy
that had been achieved in the 1980s. However, modern observations have shown
that the distance to the last scattering surface at redshift $z=1090$
is a factor of $1.5$ to $2.0$ larger than predicted
\cite{Vonlanthen:2010, Hubble}
(if the Hubble constant is held fixed and primordial perturbations
have a power-law spectrum).
Observations of type Ia supernovae and large-scale structure
\cite{SNIa, kinematic, BAO}
show that this discrepancy occurs at redshifts of order
unity, that is, at approximately 10 billion years.
Most observations probe distances, and the expansion
rate is usually inferred through the use of the FLRW relation between
the distance and the expansion rate (see \sec{sec:light}).
In the FLRW model, the explanation for longer distances is that
the expansion has accelerated, so objects have been pushed
further away than expected.

There are few model-independent observations of the expansion
rate as a function of redshift, but the Hubble parameter
today is relatively well known \cite{Hubble}.
By combining with measurements of the matter density we obtain
$\Omega_\mathrm{m0}\equiv8\pi\GN\rho_\mathrm{m0}/(3H_0^2)\approx0.2\ldots0.4$
\cite{Vonlanthen:2010, Peebles:2004}.
Combining it with the age of the universe instead \cite{Krauss:2003},
we get $H_0t_0\approx0.8\ldots1.1$.
In the SCDM model, $\Omega_\mathrm{m0}=1$ and $H_0t_0=2/3$, so either
way we find that the expansion is faster than expected by a factor
ranging from $1.2$ to $2.2$.
This finding supports the interpretation of the distance
observations in terms of faster expansion.
At the moment, model-independently it is possible to say only
that the expansion has decelerated less, not that it would
have accelerated \cite{ages}.
Analysis of large scale structure \cite{BAOradial},
especially in combination with supernova data \cite{Blake:2011}
can provide strong constraints on the evolution of the
Hubble parameter, though these constraints depend on
the assumption that the universe is close to FLRW.
Currently, we cannot conclude that the expansion would have
accelerated without assuming the validity of the FLRW approximation.

Within the FLRW model, the conclusion that the expansion rate
has accelerated follows directly from kinematical analysis
of distance observations \cite{kinematic}.
To obtain acceleration in the FLRW model it is necessary to modify
the dynamics either by introducing some exotic form of matter that has
negative pressure or by changing the laws of gravitation on cosmological
scales. In the simplest such model, the $\Lambda$CDM model,
constant vacuum energy or, equivalently, a cosmological constant
is added. The only difference between the $\Lambda$CDM model
and the SCDM model is that in the former the
expansion rate increases at late times, which is sufficient
to bring predictions and observations into agreement (although
problems appear to remain with structures on large scales
\cite{inhom, Einasto, Murphy:2010, Yaryura:2010, Nadathur:2011}).
This point is worth emphasizing: anything that changes the
expansion rate and the corresponding distances at late times by the
right amount (and that does not modify other things too much)
accounts for all of the observed discrepancies,
whether of luminosities of supernovae, anisotropies
of the cosmic microwave background (CMB), the growth rate of structures
or other probes. Notably, all the relevant observations
involve integrals over long distances or averages over large scales.

\subsection{The hypotheses behind the FLRW model}

The common justification for using the FLRW model to describe the
universe on average is that the universe appears to be homogeneous and
isotropic on large scales.
However, we have to distinguish between {\it exact}
and {\it statistical} homogeneity and isotropy.
Exact homogeneity and isotropy means that the spatial geometry
has a local symmetry: all points and directions are equivalent.
(The topological structure may break this symmetry globally.)
Statistical homogeneity and isotropy means that if we
consider a large domain anywhere in the universe, the mean quantities
in the domain do not depend on its location, orientation or size,
provided that the domain is larger than the {\em homogeneity
scale}\footnote{Assuming
that a homogeneity scale exists, that is, that the mean density and
other average quantities converge to a scale-independent value
as the volume increases. We make this assumption throughout.
In statistical physics, this property is known as
{\it spatial homogeneity}  -- not to be confused with
the GR property of the same name which is related to local
symmetry \cite{inhom}.}.

In the usual cosmological treatment, the early universe is assumed to
be close to exact homogeneity and isotropy in the sense that the
difference between the energy density, expansion rate and other
quantities between any two spatial locations is small.
Cosmological observations, especially the CMB, provide
strong support for this assumption\footnote{However,
it has not been proven and in principle the effects we discuss
could also be relevant in the early universe.}, and we retain it.
Nevertheless, when density perturbations become nonlinear at late times,
the universe is no longer locally close to homogeneity and isotropy.
However, it remains statistically homogeneous and isotropic 
on large scales, assuming the initial distribution of perturbations
had this property.
Statistical homogeneity and isotropy (and a homogeneity scale)
are predictions of the simplest inflationary models.
We assume that these assumptions hold in the real universe,
and that the homogeneity scale is of order 100 Mpc.
Whether the existence of a homogeneity scale has been observationally
established is a matter of contention \cite{Hogg:2005, inhom},
and studies of the morphology of structures suggest that such
a scale is no smaller than 300 Mpc \cite{fluctuations}.

Two concepts that are sometimes used to motivate the use of the
FLRW metric to describe the universe are the {\em Copernican principle}
and the {\em cosmological principle}. According to the former, our location
in the universe is not special, whereas according to the latter all
locations in the universe are equivalent. However, the Copernican
principle says nothing about possible symmetries of the geometry
or the matter distribution. It is possible to have a space
that has preferred directions or locations, but where the
observer is not in a preferred location.
On the other hand, the cosmological principle is a statement
about spacetime symmetry, however strictly speaking it does not apply
to the real universe because the universe contains structures.
If the principle is weakened and interpreted to refer to the distribution of
structures, then it is nothing but the statement of statistical
homogeneity and isotropy.
In modern cosmology, this statement is a prediction of simple models
of inflation, rather than a principle, and it is subject to observational tests.
Neither the Copernican principle nor the cosmological principle
(interpreted to refer to large-scale statistical properties)
show that the universe would be described by the FLRW model.
In the case of observational tests
\cite{Hogg:2005, inhom, Clarkson:2007b, kSZ, Maartens:2011, Heavens:2011, Clifton:2011},
it is important to distinguish whether they probe statistical
homogeneity and isotropy, the FLRW metric or the Copernican
principle. (For example, the important check proposed in
\cite{Clarkson:2007b} tests the FLRW metric, not the Copernican principle.)

Put simply, the FLRW model describes universes that are
locally homogeneous and isotropic on all scales, not universes
that are only statistically homogeneous and isotropic.
Because there are large local deviations, the average evolution may
be far from the FLRW behavior even above the homogeneity scale.
The possibility that the observed change in average quantities
from those of the SCDM model at late times is due to the
formation of structures may be termed the {\it backreaction conjecture}.

\section{From the local to the average}

\subsection{The local expansion rate}

We assume that the energy density dominates over pressure,
anisotropic stress and energy flux everywhere,
in other words, that matter can be considered a pressureless
ideal fluid, or {\em dust}. This
assumption does not hold in the real universe,
where deviations from dust evolution are important in regions
of large density contrast \cite{adhesive}.
However, it seems likely that the effect on average quantities is
small, because the fraction of volume in such regions is small
\cite{Pueblas:2008}. For discussion of non-dust matter, see 
\cite{Buchert:2001,Rasanen:2009b, Larena:2009, Marozzi}.
In GR, the relation between the matter and
the geometry is given by the Einstein equation:
\bea \label{Einstein}
  G_{\a\b} &=& 8\pi\GN T_{\a\b} = 8\pi\GN \rho u_{\a} u_{\b} \ ,
\eea

\noindent where $G_{\alpha\beta}$ is the Einstein tensor, $\GN$ is
Newton's constant, $T_{\alpha\beta}$ is the energy--momentum tensor,
$\rho$ is the energy density and $u^\a$ is the four-velocity
of observers comoving with the dust. The gradient of $u_\a$ can
be decomposed as
\bea \label{gradu}
  \nabla_\b u_\a
  &=& \frac{1}{3} h_{\a\b} \Theta + \sigma_{\a\b} + \omega_{\a\b} \ ,
\eea

\noindent where $h_{\a\b}$ projects orthogonally to $u^\a$,
i.e. onto the dust rest frame.
The trace $\Theta\equiv\nabla_\a u^\a$ is the volume expansion rate,
the traceless symmetric part $\sigma_{\a\b}$ is the shear tensor
and the antisymmetric part $\omega_{\a\b}$ is the vorticity tensor 
(see e.g. \cite{Tsagas:2007, Magni:2012}). 
In the FLRW case, the volume expansion rate is $3 H$, where
$H$ is the Hubble parameter, and the shear and the vorticity vanish.

The equations can be be decomposed into scalar, vector and tensor
parts with respect to the spatial directions orthogonal to $u^\a$.
We need only the following scalar parts:
\bea
  \label{Rayloc} \dot{\Theta} + \frac{1}{3} \Theta^2 &=& - 4 \pi \GN \rho - 2 \sigma^2 + 2 \omega^2 \\
  \label{Hamloc} \frac{1}{3} \Theta^2 &=& 8 \pi \GN \rho - \frac{1}{2} \sR + \sigma^2 - \omega^2 \\
  \label{consloc} \rhodot + \Theta\rho &=& 0 \ ,
\eea

\noindent where the dot stands for a derivative with respect to
the proper time $t$ measured by observers comoving with the dust; 
$\sigma^2\equiv\ha\sigma^{\alpha\beta}\sigma_{\alpha\beta}\geq0$
and $\omega^2\equiv\ha\omega^{\alpha\beta}\omega_{\alpha\beta}\geq0$
are the shear scalar and the vorticity scalar, 
respectively. In the irrotational case, $\sR$ is the
scalar curvature of the hypersurface which is orthogonal
to $u^\a$; see \cite{Ellis:1990} for the definition
of this term in the case of non-vanishing vorticity.

Equation \re{consloc} shows simply that the energy density is
proportional to the inverse of the volume, in other words
that mass is conserved. The Hamiltonian constraint \re{Hamloc}
is the local equivalent of the Friedmann equation for
an arbitrary dust spacetime, and it relates the expansion
rate to the energy density, spatial curvature, shear and vorticity.
The Raychaudhuri equation \re{Rayloc} gives the local acceleration.
We assume that the fluid is irrotational, i.e. that the vorticity
is zero. As with the assumption that the matter is dust, the
irrotationality assumption is violated in the real universe,
but the violation is not expected to change the overall picture
because vorticity is expected to be large only in a small fraction of space.
See \cite{Rasanen:2009b} for the case in which the vorticity is non-zero.
Because vorticity contributes positively to the acceleration,
setting it to zero gives a lower bound.
In this case, the local acceleration is always negative,
or at most zero, which simply expresses the fact that gravitation
is attractive for matter that satisfies the strong energy condition,
which here reduces to $\rho\ge0$.

\subsection{The average expansion rate} \label{sec:av}

When discussing averages, the first question
concerns the choice of the hypersurface on which the average is taken.
We choose the hypersurface orthogonal to $u^\a$, which is also
the hypersurface of constant proper time $t$ measured by the observers;
we discuss this choice in \sec{sec:light}.
The spatial average of a scalar quantity $\psi$ is its
Riemannian volume integral over a compact domain $\CD$
on the hypersurface, divided by the volume of the domain:
\bea \label{av}
  \av{\psi}(t) \equiv \frac{ \int_\CD d^3 X \sqrt{^{(3)}g(t,X^i)} \, \psi (t,X^i) }{ \int_\CD d^3 X \sqrt{^{(3)}g(t,X^i)} } \ ,
\eea

\noindent where $^{(3)}g$ is the determinant of the metric on the
hypersurface of constant proper time $t$, and $X^i$ are Gaussian
normal coordinates that are constant along geodesics of the dust flow.

Averaging \re{Rayloc}--\re{consloc}, we obtain the following
well-known set of equations \cite{Buchert:1999a}\footnote{
These equations are the simplest example of the averaged equations known in the literature as the ``Buchert equations''.}:
\bea
  \label{Ray} 3 \frac{\ddot a_\CD}{a_\CD} &=& - 4 \pi \GN \av{\rho} + \sQ \\
  \label{Ham} 3 \HH &=& 8 \pi \GN \av{\rho} - \frac{1}{2}\av{\sR} - \frac{1}{2}\sQ \\
  \label{integ} \pat_t \av{\rho} + 3 \frac{{\dot a}_\CD}{a_\CD} \av{\rho} = 0 \;\;;\;\;&&
a_{\CD}^{-2} \pat_t  ( a_{\CD}^{2}\av{\sR}) \;+\;a_{\CD}^{-6}\pat_t ( a_{\CD}^{6}{\sQ} )\;=0\;,
\eea

\noindent where the {\em kinematical backreaction variable} $\sQ (t)$ contains the effect of inhomogeneity and anisotropy,
\bea \label{Q}
  \sQ \equiv \frac{2}{3}\left( \av{\Theta^2} - \av{\Theta}^2 \right) - 2 \av{\sigma^2} \ ,
\eea

\noindent and the scale factor $a_\CD (t)$ is defined so that the
volume of the spatial domain is proportional to $a_\CD(t)^3$,
\bea \label{a}
  a_\CD (t) \equiv \left( \frac{ \int_\CD d^3 X \sqrt{ ^{(3)}g(t,{X^i})} }{ \int_\CD d^3 X \sqrt{ ^{(3)}g(t_0,{X^i})} } \right)^{\frac{1}{3}} \ ,
\eea

\noindent where $a_\CD$ has been normalized to unity at time $t_0$.
Because $\Theta$ gives the volume expansion rate,
this definition of $a_\CD$ is equivalent to the definition
$3{\dot a}_\CD / a_\CD \equiv\av{\Theta}$.
We also use the notation $H_\CD \equiv \dot a_\CD / a_\CD$.

The integrability conditions (\ref{integ}) assure that the
volume expansion law is the integral of the volume acceleration law.
Whereas the mass conservation law for dust is sufficient in the
homogeneous case, in the inhomogeneous case there is a further
equation that dynamically relates the averaged intrinsic and
extrinsic curvature invariants.
Note that in the presence of inhomogeneities the
averaged scalar curvature does not obey a separate conservation law
as the density does (unlike in FLRW cosmology): only a combination of
the average spatial curvature and the kinematical backreaction variable
is conserved \cite{buchert:review}.

The set of averaged equations \re{Ray}--\re{integ} has a slightly different
physical interpretation than the FLRW equations due to the different
meaning of the scale factor.
In the FLRW model, the scale factor is a component of the metric,
and indicates how the space evolves locally. (The evolution
of any finite volume is of course given by the same scale factor.)
In the present context, $a_\CD (t)$ does not describe local
behavior, and it is not part of the metric:
it gives only the total volume of the region over which
the average is taken.

The average kinematical equations above can be written in a
form reminiscent of the FLRW model, with the correction
terms included as sources \cite{Buchert:2001}:
\bea
  3\frac{{\ddot a}_{\CD}}{a_{\CD}} &=&  - 4\pi \GN (\varrho_{\rm eff}^{\CD}+3{p}_{\rm eff}^{\CD}) \\
  3\frac{{\dot a}_\CD^2}{a_\CD^2} &=&  8\pi \GN \varrho_{\rm eff}^{\CD} \\
&&{\dot{\varrho}}_{\rm eff}^{\CD}+3\frac{{\dot a}_\CD}{a_\CD}(\varrho_{\rm eff}^{\CD}+{p}_{\rm eff}^{\CD})=0\ ,
\label{eq:effectivefriedmann}
\eea

\noindent where the effective energy density and effective
pressure are defined as
\bea
\varrho_{{\rm eff}}^{{\CD}} :=& \average{\varrho} + \varrho_{\Phi} \qquad& \varrho_{\Phi}  :=  -\frac{1}{16\pi \GN}{\sQ}-\frac{1}{16\pi \GN}\av{\sR} 
\label{eq:equationofstate}\\
{p}_{{\rm eff}}^{{\CD}} :=& p_{\Phi} \qquad& p_{\Phi} :=  -\frac{1}{16\pi \GN}{\sQ}+\frac{1}{48\pi \GN}\av{\sR} \ ,
\eea

\noindent respectively. This form of the equations suggests interpreting the
new sources due to the curvature inhomogeneities in terms of a minimally
coupled scalar field \cite{Buchert:2000,buchert:static,morphon},
which provides a different language in which to talk about inhomogeneities
in geometrical variables. However, the effective sources can have more 
general behavior than a minimally coupled scalar field.
The `effective kinetic energy' is not restricted to be positive-definite,
and the effective equation of state can correspondingly evolve
from larger than $-1$ to smaller than $-1$.

There has been much discussion about the possible gauge-dependence
of the average expansion rate
\cite{Ishibashi:2005, foliation, Marozzi, Larena:2009, Rasanen:2010a}.
It is useful to distinguish three different concepts:
gauge-dependence, coordinate dependence and dependence on the
choice of the averaging hypersurface.
Gauge choice refers to the mapping between a fictitious background
spacetime and the real spacetime.
In the covariant formalism, there is no split into background
plus perturbations, so the issue of gauge-dependence does not arise.
The averaging procedure expressed in the covariant formalism
is also independent of the choice of coordinates.
The result does, however, depend on the hypersurface on which
the average is taken \cite{Ishibashi:2005,foliation, Marozzi, Rasanen:2010a};
note that this is a physical choice, not a matter of mathematical description.
We discuss the choice of hypersurface in \sec{sec:light}.

\subsection{Physical meaning of average acceleration}

If the variance of the expansion rate is large enough
compared with the shear and the energy density, then
the average expansion rate accelerates according to
\re{Ray} and \re{Q}, even though \re{Rayloc}
shows that the local expansion rate decelerates everywhere.
This feature is related to the fact that time evolution
and averaging do not commute, and it may seem somewhat counterintuitive.
However, the explanation is rather simple.
Because the universe is inhomogeneous, different regions
expand at different rates. Regions with faster expansion rate
increase their volume more rapidly, by definition.
Therefore the fraction of volume in faster expanding regions
rises, so the average expansion rate can rise.
This effect is missing in the FLRW model. 
(Whether the average expansion rate actually does rise depends on
how rapidly the fraction of faster regions grows relative to the
rate at which their expansion rate decelerates.)

To avoid confusion, it is useful to distinguish three
different concepts of acceleration.
{\em Local acceleration} means that the expansion rate
of the local volume element increases.
{\em Average acceleration} means that the rate of growth
of the total volume of the region under consideration
increases.
Finally, {\em apparent acceleration} refers
to the situation in which observations (typically distance
observations) are analyzed by fitting to a FLRW model
and the expansion rate of the fitting model accelerates.
In the case of the FLRW universe, these three accelerations
coincide due to the high symmetry.
As discussed in \sec{sec:obs}, present observations
provide strong support for apparent acceleration,
whereas average acceleration has not been established
in a model-independent manner.
(In models where we are located at the centre of a spherically
symmetric inhomogeneity, apparent acceleration typically does
not correspond to local acceleration or average acceleration
\cite{bolejkoandersson, Enqvist:2006, Bull:2012}.)
If backreaction is significant in the real universe, the average
expansion has to decelerate less than in the SCDM model, but it is 
not clear whether it actually needs to accelerate, given that if deviations
have a large effect on the average expansion rate, they
generally also significantly modify its relation to the distance.
We return to this topic in \sec{sec:light}.

\subsection{Spatial curvature, entropy and topology} \label{sec:curva}

It is possible to understand the backreaction effect in terms of the
variance of the expansion rate and the backreaction
parameter $\sQ$, or alternatively in terms of the average
spatial scalar curvature $\av{\sR}$.
The key relation is the integrability condition \re{integ}
which relates the backreaction variable and the average spatial
curvature.
The coupling demonstrates the connection between structures
that populate the universe and its spatial geometry: not only
does the extrinsic curvature (whose trace gives the expansion rate)
change but also the intrinsic curvature evolves during structure formation.
In the FLRW case, there are no structures, backreaction vanishes
and the spatial curvature term in (\ref{integ}) is conserved
individually. In this case the spatial curvature evolves as
$\av{\sR} \propto a^{-2}$ with the local scale factor $a(t)$.

A non-vanishing backreaction term $\sQ$ can lead, due to its coupling
to the spatial curvature, to a global gravitational instability of
the FLRW model. In other words, it is possible that perturbations of
the average spatial curvature and the average expansion rate grow
so that averages are driven away from the FLRW behavior, as occurs
in some scaling models for the average evolution \cite{roy:phasespace}.
In the case of the density distribution, over- and underdense regions
add up to a prescribed average density as a result of mass conservation.
By contrast, contributions of positively and negatively curved
regions to the average spatial curvature do not in general cancel,
because there is no isolated conservation law for the spatial curvature
\cite{buchertcarfora:curvature}. It might be expected that in a universe
that becomes dominated by voids the average spatial curvature would
evolve to become negative, even if it almost vanishes at early times.
Whether the spatial curvature actually evolves in this manner
has to be resolved in a realistic model; in the FLRW model, even the
possibility of such evolution is absent.

The difference between any given model and the homogeneous
and isotropic FLRW model
is measurable through the (domain-dependent) Kullback--Leibler
distance ${\cal S}_\CD$, an information theoretical {\em relative entropy}
that arises naturally from the non-commutativity of averaging and
time-evolution and is likely to increase in the course of structure formation
\cite{entropy}:
\be
\label{commutationentropy}
\pat_t \av{\rho}-\av{\pat_t \rho}\; =\; -\frac{\pat_t \,{\cal S}_\CD}{V_{\CD}}\quad;\quad {\cal S}_\CD : \;=\; 
\int_{\CD} \rho \ln \frac{\rho}{\av{\rho}}\,\sqrt{^{(3)} g} d^3 X \;\; ,
\ee

\noindent where $V_{\CD}$ is the volume of the domain.
In the FLRW model the relative entropy is zero,
because the distribution is completely unstructured.

Intrinsic curvature is also related to the topology
of spatial sections. In a Newtonian description of the
evolution of structures on a spatially flat background,
there is no spatial curvature and there are six different
orientable space forms out of which the three-torus is usually chosen.
This corresponds to implementing periodic boundary conditions
on deviations from the prescribed background.
In GR, the situation is more involved due to intrinsic
curvature, and there are an infinite number of possible
topological space forms. 
The situation simplifies for GR in 2+1 dimensions.
According to the Gauss-Bonnet theorem, for a closed
Riemannian two-space, the volume integral of the spatial
curvature is a topological invariant related to the Euler
characteristic of the two-dimensional manifold. Therefore,
the average spatial curvature always evolves as $a_\CD^{-2}$.
Correspondingly, it follows from the $2+1$-dimensional equivalent
of the integrability condition \re{integ} that $\sQ$ is always
proportional to $a_\CD^{-4}$ and cannot lead to acceleration
\cite{Rasanen:2008a, Magni:2012}.
If the topology were also to fix the average curvature in
three dimensions, the dynamics of the average spatial curvature,
and hence backreaction, would also be constrained by the topology
of space in the real universe. However, in three dimensions
the relation between topology and curvature is more involved.
This is an ongoing area of investigation and, to a large extent,
an open discipline of mathematical physics;
there has recently been progress in that field because
of results by Perelman (see e.g. \cite{carfora:perelman}).

\subsection{Newtonian cosmology} \label{sec:newton}

The present framework can also be set up for Newtonian gravity; in
fact, the averaged equations were derived first in the Newtonian
framework (along with the adaptations to GR) \cite{Buchert:1995}.  In
Newtonian theory, spatial averaging $\av{\bf }$ 
can be defined for all tensor
fields (not only scalars, as in GR) by the Euclidean
volume integral, where $\CD$ is an Eulerian domain.
The local Raychaudhuri equation \re{Rayloc}, which governs the
acceleration, is the same in both theories, so the averaged equation
\re{Ray} is also identical. However, there are important differences. In
Newtonian theory we can introduce a diffeomorphism ${\bf f}_t$ that maps
the initial (Lagrangian) positions of fluid elements (labelled by
their Lagrangian coordinates $X^k$) to Eulerian positions at time $t$,
$x^a = f^a (X^k ,t)$, and the fluid deformation can be measured by the
gradient $\pat f^a / \pat X^k$. The volume evolution along the
trajectories of a collection of fluid elements of an Eulerian domain
$\CD$ is then considered through the volume average, $V_\CD = \int_\CD
d^3 x = \int_{\CD_i} J (X^k ,t) d^3 X$, where the domain $\CD_i$ is
comoving with the collection of fluid elements, $\CD$ is the initial
domain and $J = \det (\pat f^a / \pat X^k )$
denotes the Jacobian of the coordinate transformation
from Eulerian to Lagrangian coordinates and measures the local volume
deformation. The fluid geometry, as it is embedded into the Euclidean
vector space, is described by the Lagrangian metric with the
spatial line element
\be
ds^2 = g^E_{ij}dX^i dX^j = \delta_{ab}\frac{\pat f^a}{\pat {X^i}} \frac{\pat f^b}{\pat {X^j}} \;dX^i dX^j \ .
\ee

\noindent This metric is flat (i.e. Euclidean), because it can be reduced
to the form $\delta_{ij}$ by the inverse of the transformation
${\bf f}_t$. In GR, in contrast, the volume deformation is non-integrable.
If we introduce non-exact differential forms,
$\boldsymbol{\eta}^a \ne {\bf d} f^a$ and express them in an exact basis,
$\boldsymbol{\eta}^a = \eta^a_{\;i}dX^i$, the metric coefficients give
the Riemannian spatial line element
\be
ds^2 = g^R_{ij} dX^i dX^j = \delta_{ab}\eta^a_{\;i} \eta^b_{\;j} \;dX^i dX^j \ ,
\ee

\noindent where the former Lagrangian coordinates $X^k$ are now the
local coordinates in the tangent space at a point on the Riemannian
manifold, and the transformed volume element $J d^3 X$ is now the
Riemannian volume element with
$J = \det (\eta^a_{\;i})=\sqrt{^{(3)} g}$.
The non-existence of the embedding into a Euclidean vector
space, i.e. the non-integrability of the volume deformation,
gives rise to intrinsic curvature.
The backreaction term $\sQ$ is built in the same way in
both theories from invariants of the expansion tensor.
In the Newtonian theory, $\sQ$ can be expressed as a full
divergence because of the above integrability property.
The divergence can then be transformed
to a boundary term via Gau\ss' theorem\footnote{One consequence
is that in the Newtonian theory we can provide a morphological
interpretation of backreaction: the backreaction variable
$\sQ$ can be expressed in terms of three of the four {\em Minkowski
functionals} \cite{buchert:review} (section 3.1.2).
These measures were introduced into cosmology by Mecke
et al. \cite{minkowski} to statistically assess
morphological properties of cosmic structure \cite{fluctuations, web}.}.
In GR, by contrast, $\sQ$ does not reduce to a boundary term,
which is related to the fact that (in 3+1 or more dimensions)
the spatial curvature can have non-trivial evolution.

These differences are crucial for backreaction. When we impose
periodic boundary conditions in Euclidean space, the backreaction
variable $\sQ$ is strictly zero on the periodicity scale (a
three-torus has no boundary).
This is the construction principle for structure formation
models in cosmological simulations.
Therefore, there is no global backreaction, and $\sQ$ simply describes
cosmic variance of the peculiar-velocity gradient in the interior of the
structure simulation box\footnote{We can still compute
the backreaction term in Newtonian
simulations on scales below the periodicity scale to determine how the
evolution of average quantities on scales smaller than the size of
the box is affected by backreaction \cite{Buchert:1999b}.}.
This point is interesting in itself, because N-body simulations have
been set up without investigating backreaction.
If $\sQ$ were not a full divergence, the construction
of N-body simulations would be inconsistent.
Because in Riemannian geometry $\sQ$ is in general not a boundary
term, current cosmological simulations cannot describe a
global backreaction effect. If backreaction is substantial,
then current Newtonian simulations (as well as Newtonian
analytical studies) are inapplicable in circumstances where
deviations of the intrinsic curvature are important.

If backreaction is globally significant (and the universe is
statistically homogeneous and isotropic), this is due to
non-Newtonian aspects of gravity
\cite{Buchert:1995, Buchert:1999a, Notari:2005, Kolb:2005, Rasanen:2006b, buchert:review, Rasanen:2008a, buchertcarfora:curvature, Rasanen:2010a, Rasanen:2011b, Bertello:2012}.
It is therefore crucial to understand the relation between
Newtonian gravity and GR, which is sometimes called the Newtonian
limit of GR. In the above description of fluid deformations in GR the
Newtonian limit is obtained if the non-integrable deformation one-forms
can be reduced to integrable ones, $\boldsymbol{\eta}^a \rightarrow {\bf d}f^a$.
This limit automatically implies that the geometry is Euclidean,
in particular that the intrinsic curvature and the magnetic part
of the Weyl tensor vanish everywhere \cite{NewtonWeyl}\footnote{There
is no need to consider a quasi-Newtonian metric form such as the
conformal Newtonian gauge, as is often done, to perform this limit.
In the comoving-synchronous setting, we obtain the Lagrangian form of the
Newtonian equations \cite{ehlersbuchert} in this limit,
not the Eulerian form as in the conformal Newtonian gauge.}.
However, it is not clear to which physical circumstances the above formal limiting process corresponds to.

In cosmology (unlike in the asymptotically flat case) the Newtonian
limit is not completely understood. In other words, it is not clear
under which physical conditions Newtonian gravity gives a good
approximation to GR in an extended system with
non-linear deviations. In fact, because the Newtonian
limit is singular, it would be more accurate talk about the
small-velocity, weak field limit of GR, which is qualitatively
different from Newtonian gravity
\cite{Ellis:1971, Ehlers:1991, Ellis:1994, Senovilla:1997, vanElst:1998, Ehlers:1998, Ehlers:1999, Szekeres, Ellis:2011b, Nzioki:2011, Bertello:2012}.
For example, the Newtonian equations have solutions that are not
the limit of any GR solution\footnote{In \cite{Green:2011} the problem
is approached from the other side: assuming a Newtonian cosmological
model, what is the perturbed GR model that it approximates?
However, the question of interest here is instead the
following: given a GR model evolved from realistic initial conditions,
is there any Newtonian model that approximates it at late times,
and if so, what is it?}. Newtonian gravity can thus
give a misleading picture of gravitational evolution even in the case
of slowly moving sources whose gravitational fields are not strong.
An important qualitative difference is that in the Newtonian
case there exists a global conserved quantity, the total energy,
regardless of whether the system is homogeneous or inhomogeneous
(as long as it is isolated, i.e. boundary terms can be neglected).
That inhomogeneities cannot change the global average
expansion rate in Newtonian cosmology can be understood
in terms of this constraint.
In GR, in the FLRW case, there is a conserved quantity corresponding
to the Newtonian energy term, namely the spatial curvature (multiplied
by the square of the scale factor). However, in contrast to the
Newtonian theory, this conservation law is only due to the FLRW symmetry.
If the universe is anisotropic and/or inhomogeneous, there is
in general no conservation law for the spatial curvature.
GR in 2+1 dimensions is closer to the Newtonian case in the
sense that such a conservation law does exist, as discussed in \sec{sec:curva}.

On a related topic, let us briefly discuss the behavior of
spherically symmetric dust distributions in Newtonian gravity
and GR. In GR, Birkhoff's theorem states that spherically symmetric
vacuum solutions are also static or homogeneous or both (the
theorem can be generalized to the almost spherically symmetric
case \cite{Goswami:2011}), and that the only static solution is the
Schwarzschild solution. Birkhoff's theorem has no relevance
when the boundary of the region under consideration is not in vacuum.
In the case of dust matter, there are two interesting
results. First, both in Newtonian gravity and in GR, the evolution of
a spherically symmetric dust region depends only on the distribution
of matter inside the region. Second, in the Newtonian case, the average
evolution is the same as that of a FLRW dust universe.
Both of these two statements are sometimes confused with Birkhoff's
theorem, which is unfortunate especially as the second
result does not hold in GR, except when the spatial curvature
vanishes or is strongly restricted (see \cite{buchertFOCUS}, section 7.2).
From the backreaction point of view, the Newtonian result
follows from the fact that $\sQ$ is a boundary
term that vanishes for spherical symmetry \cite{Buchert:1999b}.
In GR, this is not the case, and the average evolution
can be different from the FLRW model. Spherically symmetric models
have been used to demonstrate accelerating expansion due to
backreaction \cite{LTBacc}.

\section{Perturbation theory and non-perturbative models}

\subsection{Perturbative arguments}

It has been argued that backreaction is small
because the universe remains close to FLRW in the
sense that metric perturbations are small even when the
density perturbation becomes nonlinear.
It is useful to separate two distinct parts of this argument.
The first is that the metric perturbations around a FLRW
spacetime are small at all relevant times. The second is
that smallness of metric perturbations implies that the
average evolution remains close to the FLRW case.

Let us first assume that metric perturbations indeed
remain small. From this assumption alone we cannot conclude
that backreaction would be small, because the perturbations
of the expansion rate (and the distance) depend on first and second
derivatives of the metric perturbations, the latter of
which become large at the same time as the density perturbations.
In other words, perturbations of the spacetime curvature are large.
Indeed, in the real universe there are deviations of
order unity in the local expansion rate and in the spatial curvature
(see \cite{estimate} for estimates of the spatial curvature in
astrophysical systems of different sizes).
These are not coordinate artifacts, but rather deviations in physically
measurable quantities, and any realistic model has to account for them.
The local variation is thus of the same size as
the observed average deviation from the SCDM model: the issue is
whether their distribution is such that positive and
negative deviations cancel on average.
This question is related to the absence of a conservation law for
the average spatial curvature.

Estimates in first order perturbation theory
\cite{Futamase:1988, Seljak, Wetterich:2001, Rasanen, Siegel:2005, Ishibashi:2005}
cannot resolve the issue, because the average of the first order
perturbation vanishes, and the contribution of the square of
the first order perturbations is not gauge-invariant without the contribution
of the intrinsic second order perturbations\footnote{The separation
between the square of first order perturbations and intrinsic second
order perturbations is gauge-dependent \cite{Kolb:2004a}.}.

At second order, the effect on the average expansion rate is small
\cite{Kolb:2004a}. However, when density perturbations become
nonlinear, contributions from higher order terms become as
important as the second order terms \cite{Kolb:2005, Rasanen:2010a}.
Nevertheless, the average expansion
rate remains close to the FLRW case if there is
a coordinate system in which the following conditions hold:
(i) the metric perturbations and their first derivatives are small,
(ii) the time derivatives of metric perturbations are at most of
the same order as the perturbations multiplied by the
background Hubble parameter,
(iii) the perturbation of the observer four-velocity is small, and
(iv) the difference between the observer four-velocity and the
four-velocity that defines the averaging hypersurface is
non-relativistic \cite{Rasanen:2011b}.
Essentially, because perturbations of the Christoffel symbols
remain small in this coordinate system, the spatial structure
remains close to linear theory, even though variations in the
spacetime curvature are large.
Therefore, large deviations cancel in the average, and such models
may be termed 'quasi-Newtonian', because in Newtonian theory
cancellation of deviations in the expansion rate is built in,
as discussed above.
Under these assumptions, the average spatial curvature and the redshift
also remain close to the FLRW behavior.
However, deviations do not cancel in all quantities.
The shear and the variance of the expansion rate become large at the
same time as density perturbations, though their contributions to the
expansion rate cancel.
The average acceleration also deviates much from the background
behaviour \cite{Clarkson:2011a}. This follows from the commutation rule
$\av{\dot\theta}=\pat_t\av{\theta} - (\av{\theta^2}-\av{\theta}^2)$
and the feature that the average expansion rate is close to
the background while the variance is large.
(The commutation rule is a straightforward consequence of \re{av}.)
An important quantity in which deviations do not in
general cancel is the luminosity distance.
There is a counter-example in which all of the above assumptions are
satisfied in a perturbed SCDM model, but the luminosity distance 
receives large corrections so that it is identical to that of the
$\Lambda$CDM model with $\Omega_{\Lambda0}=0.7$ \cite{Enqvist:2009}.

It is not obvious whether the above assumptions are
satisfied in the real universe, namely whether the metric
remains close to the same global FLRW background everywhere
\cite{Ellis:2011a, Rasanen:2011b}. For example, under the above
assumptions the density (assuming a spatially flat background)
is $\rho=3 H^2 + 2 a^{-2} \nabla^2\Psi$, where $H$
is the background Hubble rate, $a$ is the background scale
factor and $-2\Psi a^2\delta_{ij}$ is the scalar perturbation
of the spatial part of the metric. For a dust background,
$H^2\propto a^{-3}$. If we consider a stabilized
region with constant density, $\Psi$ has a decaying
part $\propto a^{-1}$ and a growing part $\propto a^2$.
In this situation, perturbations of the local metric with respect
to a global background eventually become large.
However, perturbations becoming large does not imply that the
gravitational fields would become strong, but simply that the
stabilized region becomes more and more different from the expanding
background whose density and curvature decrease.
(Once the perturbations become large, their growth could be
even faster than indicated by this argument.)
However, the situation is complicated by the fact
that it would be more appropriate to discuss the evolution
in terms of the proper time and the orthogonal spatial
coordinates instead of the background time and space coordinates
(see also \cite{buchertFOCUS}, section 7.4.2).
Once the second derivatives of the metric become large,
the coordinate transformation from one to the other is
non-perturbative, and the question of whether backreaction
is significant seems to escape the reach of standard
perturbation theory.
If it could be demonstrated that starting from realistic
initial conditions, perturbations and their first derivatives
have remained small until today, the backreaction issue would be solved
for the average expansion rate and redshift (the luminosity
distance would require more work).

An alternative to standard perturbation theory is to employ gradient
expansion techniques that allow to go substantially beyond the perturbative
regime \cite{Kolb:2005, gradientexpansion}. One further remedy could be the
formulation of perturbation theory in the actually curved space sections by
employing nonintegrable deformations in the local tangent space
\cite{buchertFOCUS, RZA1}.
However, even in such a perturbation framework, a global
background has to be specified, an issue that might be overcome by the
formulation of perturbation theory around the average distribution rather
than the background. A first step towards such a theory is presented
in \cite{roy:generalbackground}.

Constructions in which it is assumed that the metric remains
close to the FLRW background are thus severely limited in
to what extent it is possible to study backreaction.
For example, this is true for the new formalism for
averaging and perturbation theory presented in \cite{greenwald},
which claims that if metric perturbations around a global
background metric remain small, the average evolution can
be modified only by a possible radiation-like source term due to
the gravitational degrees of freedom. Although the assumption of
small perturbations around a global background remains to be
justified, \cite{greenwald} presents
an example of a formalism for obtaining an average geometry
(and not only average scalar quantities).

\subsection{Non-perturbative models}

In addition to perturbative arguments, backreaction has been
studied in toy models and semirealistic models that do not
rely on perturbation theory.

\subsubsection{Exact toy models}

A few toy models have demonstrated how backreaction
can lead to accelerating expansion in a dust universe
by using the exact spherically symmetric solution \cite{LTBacc}
or a model with two disjoint regions \cite{Rasanen:2006b}. 
Apart from their use as existence proofs, such models can be used
to understand qualitative features of average acceleration.

In the real universe, the early distribution of perturbations of the
density and the expansion rate is very smooth, with only
small local variations. However, as perturbations become
nonlinear, regions evolve in distinct ways.
In a simplified picture, overdense regions
slow down more as their density contrast grows, and eventually
they turn around and collapse to form stable structures.
Underdense regions become even emptier, and their
deceleration decreases.
Regions thus become more differentiated and the variance
of the expansion rate grows,
contributing positively to the volume acceleration,
as \re{Ray} shows.

In \cite{Rasanen:2006b} a toy model of this evolution
was considered, in which two spherically symmetric regions
which are taken to individually evolve according to
Newtonian gravity. One of these regions is underdense and the
other overdense, so they evolve like dust FLRW universes
with negative spatial curvature and positive spatial curvature, respectively.
There is one free parameter in the model: the relative size
of the regions at some instant of time. This parameter can
be chosen such that there is a period of accelerated average expansion.
Initially, the evolution is close to the FLRW case.
As the slower region starts to deviate significantly
from the FLRW evolution, the average expansion rate decreases.
However, the underdense region eventually comes to dominate
the volume, so the average expansion rate rises.
If the underdense region catches up quickly enough, not only can
the average expansion rate accelerate, but the average Hubble
parameter can rise. In the FLRW model such evolution would
require a violation of the null
energy condition, i.e. an equation of state more negative
than $-1$, or an equivalent modification of gravity.

As \re{Ham} and \re{Q} show, the volume acceleration is
bounded from above by the variance of the expansion rate.
The average expansion rate is also bounded from above;
estimates in terms of the maximum value of the local expansion
rate have been proposed.
For irrotational dust, the maximum is achieved in a
completely empty constant-curvature region, which implies
that $H_\CD t\le 1$ \cite{Nakamura:1995, Rasanen:2005}.
This bound is also expected to hold in realistic models,
in which vorticity and deviations from dust behavior are
not significant  in large fractions of volume \cite{Rasanen:2009b}.

The acceleration is due to the interplay between fast
and slow expanding regions. The larger the difference
between the fast and slow regions is, the more rapidly the
fast regions take over and the larger is the acceleration,
which explains why the variance in $\sQ$ contributes positively
to acceleration.
For this reason, a period in which the universe decelerates
more than in the SCDM model is conducive to acceleration.

\subsubsection{A semirealistic model}

In the case when the universe is not everywhere near to a global
FLRW background, we cannot write down an exact metric from which
to calculate the expansion rate. However, because we are interested
only in large-scale averages, rather than the local details
of every structure, a statistical description is sufficient.
To determine the average expansion rate, we
need to know which fraction of the volume of the universe
has which expansion rate at a given time.
One way to do so is to treat the universe as an ensemble
of regions in different stages of expansion or collapse,
and to have a model for the evolution of the statistical
distribution which is faithful to the physics of
structure formation.

The problem is well-defined in the sense that the initial
conditions, the matter content and the power spectrum are known.
For a Gaussian distribution, all statistical information
is contained in the power spectrum, and because structure formation
is a deterministic process, all statistical quantities at late
times can be traced back to the initial power spectrum.
The question is how to find a tractable description of the
gravitational processing of the initial density field.

A semirealistic model was constructed in
\cite{Rasanen:2008a, peakrevs}.
The starting point is that the universe at
early times can be described as a dust-dominated
spatially flat FLRW model with linear Gaussian
perturbations. The idea is to use the number
density of peaks and troughs in the initial density field,
smoothed on a given scale, as a measure of the fraction
of volume occupied by regions of a given density contrast
on that scale.
The smoothing scale is fixed by the requirement that
the root mean square of the smoothed density contrast,
$\sigma(t,R)$ is unity at all times; in other words, the smoothing
scale evolves to follow the size of typical structures.
The regions are taken to be spherical and to evolve as
in Newtonian gravity, so they behave like FLRW universes,
as in the two-region toy model.
The peak number density is known analytically
in terms of the initial power spectrum \cite{Bardeen:1986}.
The power spectrum depends on two parts, the primordial power
spectrum, determined in the early universe by inflation or some
other process, and the transfer function $T(k)$, which describes
the evolution between the primordial era and today in linear theory.
Given an (almost) scale invariant power spectrum and the
cold dark matter transfer function, the average expansion rate is
fixed.

\begin{figure}
\hfill
\begin{minipage}[t]{6cm} 
\scalebox{1.0}{\includegraphics[angle=0, clip=true, trim=0cm 0cm 0cm 0cm, width=\textwidth]{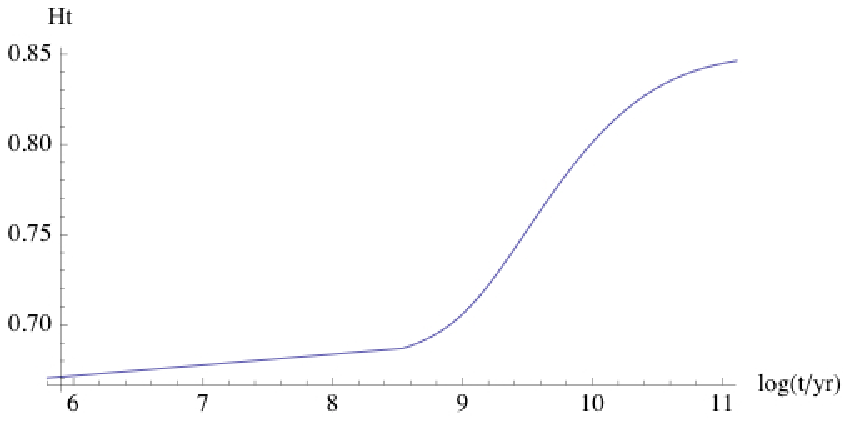}}
\begin{center} {\bf (a)} \end{center}
\end{minipage}
\hfill
\begin{minipage}[t]{6cm}
\scalebox{1.0}{\includegraphics[angle=0, clip=true, trim=0cm 0cm 0cm 0cm, width=\textwidth]{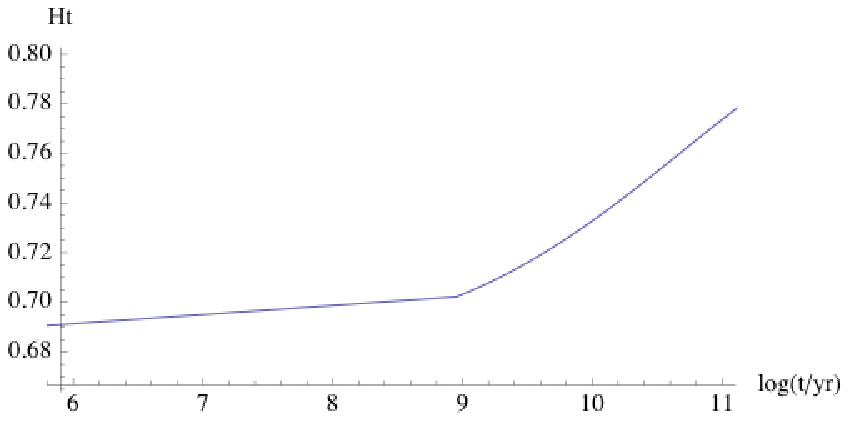}}
\begin{center} {\bf (b)} \end{center}
\end{minipage}
\hfill
\caption{The expansion rate $H_D t$ (denoted Ht) as a
function of the age of the universe in logarithmic units,
$\log(t/\mathrm{yr})$. The two curves correspond to different
approximations for the transfer function. See \cite{Rasanen:2008a} for details.}
\label{fig:Ht}
\end{figure}

Figure \ref{fig:Ht} plots the expansion rate in terms of
$H_D t$, showing significant differences from the FLRW behavior
at late times. The size of the deviation, $20\%$ to $30\%$, is of the
same order of magnitude as the observed signal, $20\%$ to $70\%$.
More remarkably, the timescale for significant change
is 10 billion years, which agrees with observations.
The reason for the rise of $H_D t$ is that underdense
regions take up more and more of the volume.
If the universe were completely dominated by totally empty
voids, we would have $H_\CD t=1$. Because the voids are
not completely empty and there are overdense regions,
$H_\CD  t$ saturates at a value somewhat smaller than unity.
(The expansion only decelerates less, the model
does not have any acceleration. The absence
of acceleration is related to the fact that
overdense regions do not appreciably slow down the expansion
rate before the underdense regions take over.)

The only scale in the problem is the matter-radiation equality
scale $\keq^{-1}\approx100$ Mpc which determines the turnover
of the CDM transfer function. Small wavelength perturbations
that enter the horizon during the radiation-dominated
era are suppressed and perturbations with wavelengths larger
than $\keq^{-1}$ retain (approximately) their original amplitude.
The combination of the corresponding matter-radiation equality
time $\teq\approx50\, 000$ yr and the amplitude
of the primordial perturbations, $A=3\times10^{-5}$,
determines when the expansion rate will change significantly.
Perturbations with wavelength equal to $\keq^{-1}$
form non-linear structures when $\sigma(t,R=\keq^{-1})=1$.
For a nearly scale-invariant spectrum, this happens when
$t\sim A^{-3/2}\teq\approx100\ldots1000$ Gyr; after that, there is no
scale in the system, so $H_\CD t$ saturates to a constant.
Transition begins somewhat earlier at $t\sim10$ Gyr, as seen in \fig{fig:Ht}.

It is interesting that the amplitude of the change in the expansion
rate as well as the timescale come out 
roughly in agreement with observations.
However, the model involves uncontrolled approximations, and
cannot be trusted beyond an order of magnitude.
It is also possible that a more careful statistical treatment
would reveal cancellations which significantly change
this approximate estimate.

\subsubsection{A generic multiscale model}

\begin{figure}[htbp]
\begin{center}
    \includegraphics[width=0.5\textwidth]{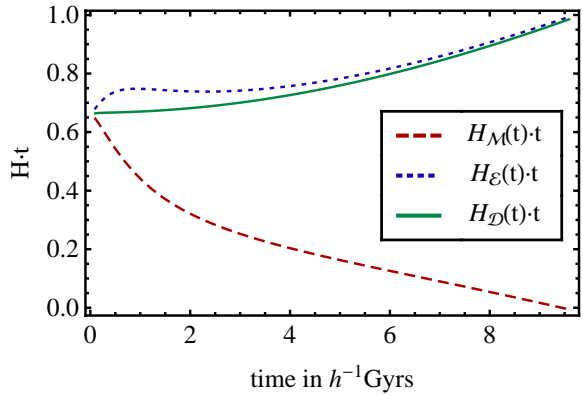} 
    \caption{The evolution of the expansion rates $H_{\cal F} t$ (denoted $H t$) averaged over regions ${\cal F} = \CD$, $\CE$ and $\CM$ are depicted for a volume-partitioned universe. (The initial conditions on $\CD$ are given in terms of a standard CDM spectrum.) On the homogeneity scale $\CD$ (full line; green) this function increases and is at late times mostly governed by the evolution in underdense regions $\CE$ (dotted line; blue) due to their volume-dominance. In the union of overdense regions $\CM$ (dashed line; red) this function decreases because mass agglomerations tend to form stationary regions that decouple from the global expansion. This evolution scenario is based on a backreaction scenario investigated in \cite{multiscale}. Figure by A. Wiegand (private communication).}
\end{center}
\label{fig:multiscale} 
\end{figure}

In the same spirit as the previous models, we can consider a general
volume partitioning of the universe by introducing a union of disjoint
overdense regions $\CM$ and a union of disjoint underdense regions
$\CE$, both of which make up the total region $\CD$, which is
considered to contain the homogeneity scale.
The averaged equations can be split accordingly to obtain for the
kinematical backreaction \cite{buchertcarfora:curvature} the result
\begin{equation}
\sQ \; =\;  \lambda_{\CM}\mQ +\left(1-\lambda_{\CM}\right)\eQ
+6\lambda_{\CM}\left(1-\lambda_{\CM}\right)\left(H_{\CM}-H_{\CE}\right)^{2}\;,
\end{equation}
where $\lambda_{\CM}$ denotes the volume fraction of the overdense regions compared with the volume of the region $\CD$. In a Gaussian random field this fraction would be $0.5$ and it would gradually drop in a typical structure formation scenario that clumps matter into small volumes and that features voids that gradually dominate the volume in the course of structure formation, see \fig{fig:multiscale}. A similar construction has been used in the so-called timescape cosmology proposed by Wiltshire \cite{wiltshire}.

\vspace{-10pt}

If we ignore, for the sake of simplicity, the individual backreaction
terms on the partitioned domains (as in the models discussed above),
the total backreaction features a positive-definite term due to
the variance of the different expansion rates of over- and underdense
regions. This term can lead to acceleration in the volume expansion
rate of the domain $\CD$ \cite{multiscale}.
Refining the multiscale model by including
non-zero individual backreaction terms, for example by an extrapolation
of the leading perturbative mode in second-order perturbation theory
\cite{lischwarz}, which also corresponds to the leading order in a
Newtonian non-perturbative model (i.e. where the backreaction terms
decay in proportion to $a_\CD^{-1}$) \cite{Buchert:1999b}, we can even
produce a cosmological constant behavior of the scale factor over the
homogeneity scale $\CD$ (see figure~3 in \cite{multiscale}).
Such behavior can only be transient, however, and the
acceleration is expected to be followed by a phase of deceleration.

\section{Light propagation} \label{sec:light}

\subsection{Connection between the expansion rate and observables}

A key question is how the spatial averages discussed thus
far are related to observations. This issue is tied to
the averages' dependence on the choice of the hypersurface
of averaging. Almost all cosmological observations are made
along the lightcone, as they measure the redshift,
the angular diameter distance (or equivalently
the luminosity distance) and other quantities
related to bundles of light rays.
In a general spacetime, these quantities are not determined
solely by expansion, and certainly not by the average
expansion rate along spacelike slices.
However, in a statistically homogeneous and isotropic universe
in which the distribution evolves slowly, the average expansion
rate may determine the leading behavior of the redshift
and the distance over large scales
\cite{Rasanen:2008b, Rasanen:2009b, Bull:2012}.
Consideration of the observables also identifies the relevant
hypersurface of averaging. In a general dust spacetime,
the redshift is given by
\bea \label{z}
  1 + z &=& \exp\left( \int_{\eta}^{\eta_0} \rmd \eta \left[ \frac{1}{3} \Theta + \sigma_{\a\b} e^\a e^\b \right] \right) \ ,
\eea

\noindent where $\eta$ is defined by
$\frac{d}{d\eta} \equiv (u^\a + e^\a) \pat_\a$, and $e^\a$
is the spatial direction of the null geodesic.
The direction $e^\a$ changes slowly for typical light rays
\cite{Rasanen:2009b}, whereas the dust shear is correlated with
the shape and orientation of structures and changes on the
length scale of those structures. If there are no preferred
directions, so structures on large scales are oriented
in all directions equally, $\sigma_{\a\b}$ should contribute via
its trace, which is zero. Therefore, the integral over
$\sigma_{\a\b} e^\a e^\b$ should vanish up to statistical
fluctuations and corrections from correlations between 
$\sigma_{\a\b}$ and $e^\a$ and the evolution of the distribution.
We can split the local expansion rate as $\Theta=\av{\Theta}+\Delta\Theta$,
where $\Delta\Theta$ is the local deviation from the average, and
similarly argue that the integral of $\Delta\Theta$ is suppressed
relative to the contribution of the average expansion rate.
The change in the distribution also has to be slow
compared with the time it takes for a light ray to
pass through a homogeneity scale sized region.
If the homogeneity scale today is of the order $100$ Mpc,
then the crossing time is indeed much smaller than
the time scale for the change in the distribution,
which is given by the Hubble scale $H_0^{-1}\approx4000\ldots5000$ Mpc.
In the early universe, structure formation was less advanced,
so the homogeneity scale relative to the Hubble scale
is even smaller further down the null geodesic.

Here the choice of hypersurface comes into play.
The relevant hypersurface is the one in which variations
around the mean cancel, in other words the hypersurface of
statistical homogeneity and isotropy.
Because the evolution of structures is governed by the proper
time, one can argue that this hypersurface is close to the
hypersurface of constant proper time
\cite{Rasanen:2006b, Rasanen:2008a, Rasanen:2008b}.
However, the two hypersurfaces are not exactly the same,
and in the realistic case when the observer four-velocity is not
irrotational, the hypersurface of constant proper time is
not orthogonal to the observer four-velocity.
The details are thus more complicated, but 
non-relativistic changes in the four-velocity field that
defines the hypersurface of averaging lead only to small changes
in the average expansion rate, as long as the distribution
is statistically homogeneous and isotropic and the averaging
scale is at least as large as the homogeneity scale \cite{Rasanen:2009b}.
However, a one-dimensional sample may converge to homogeneity
significantly slower than a three-dimensional one \cite{Bolejko}.

These cancellations also explain why the large variance required
for significant backreaction does not necessarily lead to
large deviations in the observed CMB temperature (which is
given simply by the redshift of the CMB photons).
It is sometimes claimed that if all observers measure a
nearly isotropic CMB sky, then the universe is nearly
FLRW \cite{Ellis:2011a}. However, this is not true.
Making statements about the geometry also requires
assumptions about the spatial derivatives of the
CMB temperature field, and these assumptions are not satisfied in the
real universe \cite{Rasanen:2008a, Rasanen:2008b, Rasanen:2009a}.
Observation of a nearly isotropic CMB sky does not imply that
the universe is close to FLRW.

Given that $\av{\Theta}=3{\dot a}_\CD / a_\CD$,
we obtain $1+z\approx a_\CD (t)^{-1}$,
the same relation between expansion and redshift as in the FLRW case.
This result depends on the fact that the shear and the expansion
rate enter linearly into the integral \re{z} along the null geodesic.
In contrast, the shear and the expansion rate
enter quadratically into the equations of motion \re{Rayloc}--\re{consloc}
for the geometry, so variations do not cancel in the average;
instead we have the generally non-zero backreaction variable $\sQ$.

For the angular diameter distance, similar
qualitative arguments give \cite{Rasanen:2008b}
\bea \label{DA}
  H_\CD \pat_{\bz} \left[ (1+\bz)^2 H_\CD \,\pat_{\bz} \bar{D}_A \right] &\approx& - 4\pi\GN \av{\rho} \bar{D}_A \ ,
\eea

\noindent where $\bar{D}_A$ is the dominant part of the angular diameter
distance with corrections to the mean dropped, and similarly for the
redshift: $1+\bz\equiv a_\CD (t)^{-1}$.
(It is not entirely clear that the angular deviation of the
distance must necessarily be small \cite{Clarkson:2011a}.)

From the conservation of mass \re{integ}, it follows that
$\av{\rho}\propto(1+\bz)^3$. The distance is therefore determined
entirely by the average expansion rate $H_\CD (z)$ and the normalization
of the density today, namely $\Omega_{\mathrm{m0}}$.
For a FLRW model with general matter content,
$\av{\rho}$ in \re{DA} would be replaced by $\rho+p$.
Therefore, the equation for the mean angular diameter distance in
terms of $H_\CD (z)$ in a statistically homogeneous and isotropic
dust universe (with a slowly evolving distribution) is
the same as in the FLRW $\Lambda$CDM model.
If backreaction were to produce exactly the same expansion
history as the $\Lambda$CDM model, the distance-redshift
relation would therefore also be identical.
This is true even though the spatial curvature would
be large, because the spatial curvature affects the distance
differently than in the FLRW case.

However, backreaction is not expected to produce an expansion history
identical to the $\Lambda$CDM model: if the expansion accelerates
strongly, the acceleration may be preceded by extra deceleration,
and the acceleration cannot be eternal (unless the rate of acceleration
goes asymptotically to zero sufficiently rapidly).
Therefore the backreaction distance-redshift
relation also differs from the $\Lambda$CDM model, although
it is shifted towards $\Lambda$CDM compared with a FLRW model
with the same expansion history as the backreaction model.
The reason is that in any FLRW model apart from $\Lambda$CDM,
the equation for $D_A$ is modified not only by the
mapping between the affine parameter and the redshift
(as described by $H(z)$), but also by the change of the
source term $\rho+p$, whereas in the backreaction case
only the mapping between the affine parameter and the redshift
(i.e. $H_\CD(z)$) changes.
This feature could explain why distance observations prefer
a value close to $-1$ for the effective equation of state.

Above, we assume that light is passing through
a continuous distribution of matter. However, it is not clear
whether such an assumption is valid in the real
universe, in which matter is clumped on various scales, and it
may be that a typical light ray travels in vacuum without
crossing any structures. This issue remains to be completely
understood \cite{Clifton, Clarkson:2011c, Rasanen:2009b};
for references on the effect of clumpiness on light propagation,
see \cite{Rasanen:2008b, Clarkson:2011c}.

A related problem is that light propagation is usually considered
in the geometrical optics approximation, with infinitesimal light bundles.
However, light wave fronts have a finite extent, which
is especially important if the matter distribution is discrete
on the relevant scales. Surface averaging of light fronts and its
relation to distances should thus be considered.
A covariant formalism for averaging on different slices
of the past light cone was recently introduced \cite{lightcone:venezia}.
However, because we observe both the redshift and the angular
position of sources, the relevant issue is cancellations
along the null geodesic, as discussed above.

\subsection{Observational signatures of backreaction}

In the FLRW model, there exists a definite relation between $D_A(z)$
and $H(z)$, which can be used as a general test of FLRW
models \cite{Clarkson:2007b}.
If the distance and the expansion rate are measured independently
\cite{ages, BAOradial, Blake:2011},
we can check whether they satisfy the FLRW relation.
If they do not, the observations cannot be explained
in terms of any four-dimensional FLRW model. Because the
FLRW relationship between $D_A(z)$ and $H(z)$ depends on
the spatial curvature, a test of the relation can be viewed as a
test of whether the spatial curvature is proportional to $(1+z)^2$.
If backreaction is significant, the spatial curvature
divided by $(1+z)^2$ is non-constant for some redshift range.
This holds independently of the presence of dark energy or
modified gravity, because light propagation directly depends
on the geometry of spacetime, regardless of the equations of
motion that determine it.
The consequences of a different curvature evolution have been analyzed in 
\cite{obstest, multiscale}.

The backreaction conjecture that the
change in the average expansion rate at small redshift is
due to structure formation can be tested without a
prediction for the change in the expansion rate,
simply by checking whether the measured $D_A(z)$
and $H_\CD(z)$ satisfy \re{DA}.
This relation, which violates the FLRW consistency
condition between expansion and distance, is a unique prediction
of backreaction that distinguishes it from the FLRW model.
However, the derivation of the relation should be done more
rigorously, and the expected magnitude and shape of the violation
are not known.

The redshift, as well as null geodesic shear and deflection
\cite{Rasanen:2009b}, should be studied in more detail.
In particular, it would be interesting to
quantitatively check that light propagation in a statistically
homogeneous and isotropic space with a slowly evolving
distribution of small structures can be described in terms
of the average expansion rate, and to characterize
the corrections \cite{Rasanen:2008a, Rasanen:2008b, Rasanen:2009b}.
The small-scale pattern of the CMB depends only on the
angular diameter distance \cite{Vonlanthen:2010},
but the effects on large angular scales remain to be determined.
Extending weak lensing analysis to the case in which
the geometry is not nearly FLRW is also needed for comparison
with present and upcoming data.

\section{Conclusions}

We discuss the {\it backreaction conjecture}, namely
the possibility that structure formation changes the average
expansion rate, spatial curvature and light propagation,
thereby eliminating the need for dark energy or modified gravity.
The change in the average properties due to structure formation
is present in reality but is not taken into account either in the FLRW
model and its linear perturbations or Newtonian non-linear models.

The basic mechanism of backreaction is simple and can be demonstrated
in toy models: because the universe
is inhomogeneous, different regions expand at different rates,
so the fraction of volume in faster expanding regions can grow,
and the average expansion rate can rise. Structure formation
has a preferred time of 10 billion years, which agrees with the
observed timescale for the change in the expansion rate, and the
amplitude of the change can also be understood from simple considerations.
However, the effect has not been quantified in a fully realistic model.

Various things remain to be done before robust quantitative
conclusions can be drawn. If backreaction is significant,
it cannot be understood simply as a change in the FLRW background
model. To analyze signals such as baryon acoustic
oscillations and other features in the distribution of large
structure, we must develop perturbation theory
around a non-FLRW background, where the mean is the average
of deviations that have a large amplitude, but a small coherence
length. 

The treatment of light propagation also needs to be made more
rigorous and extended both to cover phenomena such as weak lensing in
detail and to include effects due to the discreteness of matter.
The deviation of the relation between distance and the
expansion rate from the FLRW case is an important prediction,
which can be tested without a calculation of the average
expansion rate.
However, the goal should be to derive the change
in the average expansion rate with quantified errors.
In addition to statistical models, one way to address the
question could be to generalize N-body simulations to include
the relevant relativistic degrees of freedom.

The backreaction conjecture is conservative in the sense that
it does not involve new fundamental physics, only neglected effects
in non-linear GR. Before the effect of structure
formation on the average expansion rate is reliably quantified,
we will not know whether dark energy or modified gravity is needed.
It is plausible that backreaction can explain all of the observations,
and even if the large-scale average properties of the universe
turn out to be close to the FLRW case, the corrections
may nevertheless be quantitatively important.

\acknowledgments

The work of TB was conducted within the
``Lyon Institute of Origins'' under grant ANR-10-LABX-66.

\setcounter{secnumdepth}{0}

\end{document}